\documentclass[prl,aps,twocolumn,10pt,showpacs,groupedaddress,superscriptaddress,floatfix,notitlepage]{revtex4-1}
\usepackage{amsmath,amsfonts,amssymb,graphics,graphicx,epsfig,color,times}
\usepackage[utf8x]{inputenc}
\usepackage{color}
\usepackage{bbm, dsfont} 
\usepackage{subfigure}
\usepackage{hyperref}
\usepackage{mathrsfs}
\usepackage{verbatim}
\begin{document}
\title{Bound and Subradiant Multi-Atom Excitations in an Atomic Array with Nonreciprocal Couplings}
\author{H. H. Jen}
\email{sappyjen@gmail.com}
\affiliation{Institute of Atomic and Molecular Sciences, Academia Sinica, Taipei 10617, Taiwan}

\date{\today}
\renewcommand{\r}{\mathbf{r}}
\newcommand{\f}{\mathbf{f}}
\renewcommand{\k}{\mathbf{k}}
\def\p{\mathbf{p}}
\def\q{\mathbf{q}}
\def\bea{\begin{eqnarray}}
\def\eea{\end{eqnarray}}
\def\ba{\begin{array}}
\def\ea{\end{array}}
\def\bdm{\begin{displaymath}}
\def\edm{\end{displaymath}}
\def\red{\color{red}}
\pacs{}
\begin{abstract}
Collective decays of multiply-excited atoms become subradiant and bound in space when they are strongly coupled to the guided modes in an atom-waveguide interface. In this interface, we analyze their average density-density and modified third-order correlations via Kubo cumulant expansions, which can arise and sustain for long time. The shape-preserving dimers and trimers of atomic excitations emerge in the most subradiant coupling regime of light-induced dipole-dipole interactions. This leads to a potential application of quantum information processing and quantum storage in the encoded nonreciprocal spin diffusion, where its diffusion speed depends on the initial coherence between the excited atoms and is robust to their relative phase fluctuations. The state-dependent photon routing can be viable as well in this interface.        
\end{abstract}
\maketitle
{\it Introduction.}--Quantum correlation features the essence of quantum mechanical systems and distinguishes them from the classical world \cite{Zurek1991}. This nontrivial correlation provides the essential resource with which quantum information processing \cite{Hammerer2010} and quantum computation \cite{DiVincenzo2000} gain supremacy over their classical counterparts. Atom-waveguide interface \cite{Chang2018} presents such potential of superiority, which has been proposed to create mesoscopic entanglement \cite{Tudela2013} and quantum spin dimers \cite{Ramos2014, Pichler2015}. It has also been predicted to manifest strong photon-photon correlations \cite{Mahmoodian2018}, and only recently single photon storage and retrieval \cite{Corzo2019} is realized in an atomic array coupled to a nanofiber. Owing to the guided modes of the waveguide, the long-range dipole-dipole interactions emerge in light-matter couplings \cite{Kien2005}, which leads to superradiant emissions even from two distant atom clouds \cite{Solano2017}. This collective light-matter coupling is also responsible for the subradiant emissions \cite{Albrecht2019, Jen2020_subradiance} with a lifetime longer than the intrinsic one, and an enhanced performance of photon storage can be achieved via the tailored collective states \cite{Garcia2017}.

The light-matter coupling can be engineered under external magnetic fields to be nonreciprocal with controlled fractions of left- to right-propagating decay rates \cite{Mitsch2014}, which breaks the time-reversal symmetry \cite{Bliokh2014, Bliokh2015} that is preserved in most free-space quantum optical settings. Equipped with nonreciprocal or unidirectional light couplings \cite{Gardiner1993, Carmichael1993, Stannigel2012}, chiral quantum optics \cite{Lodahl2017} offers many opportunities in generating path-encoded photons \cite{Luxmoore2013}, simulating Mach-Zehnder interferometer to realize single-photon diodes and circulators \cite{Sollner2015}, creating localized edge or delocalized hole excitations \cite{Jen2020_PRR}, and renewing the study of disorder-assisted single excitation localization under dissipations \cite{Jen2020_disorder}. 

It is interesting and challenging to investigate and generate few-photon quantum correlations. Recently, a long-lived photon pair \cite{Ke2019} has been theoretically proposed in an atomic array, two-photon and three-photon bound states are observed in a quantum nonlinear medium with the atomic Rydberg states \cite{Liang2018}, and the few-photon scattering property can be reconstructed from the statistics of light in a single quantum emitter coupled to a nanophotonic waveguide \cite{Jeannic2021}. The photonic bound states can also be observed in superconducting qubit arrays \cite{Kim2021}, making the atom-waveguide interface promising to simulate quantum many-body states of light with long-range quantum correlations. 

In this Letter, we present the subradiant and bound density-density and third-order correlations, which arise in the diffusion of atomic excitations in an atom-waveguide interface with nonreciprocal couplings. We construct a Hilbert space of multiply-excited states, under which the quantum correlations can be obtained via Kubo cumulant expansions \cite{Kubo1962, Bianucci2020}. We further propose to use the encoded excitations diffusion to convey and manipulate quantum information in this quantum interface, which is found to be robust to the phase fluctuations of the initialized states.       

\begin{figure}[b]
\centering
\includegraphics[width=8.5cm,height=6cm]{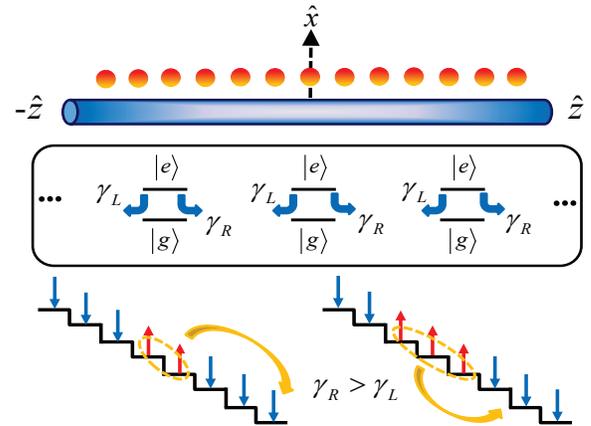}
\caption{Schematic plot of an atomic array with propagating bound atomic excitations. The atomic array of two-level quantum emitters (an effective spin $1/2$ systems of up and down representing for the excited $|e\rangle$ and ground states $|g\rangle$, respectively) with nonreciprocal couplings ($\gamma_R$ $\neq$ $\gamma_L$) transport a correlated bound dimer (in the left staircase) or trimer of atomic excitations (in the right staircase) to the right side of the array when $\gamma_R$ $>$ $\gamma_L$. A strong coupling regime can be reached when the atoms are positioned close to the waveguide and interact with the guided modes, leading to the subradiant bound multimers.}\label{fig1}
\end{figure}

{\it Model of a chirally-coupled atomic array.}--The atom-waveguide system can be schematically shown in Fig. \ref{fig1}, where we consider an array of two-level quantum emitters ($|g\rangle$ and $|e\rangle$ for the ground and excited states as an effective spin $1/2$ system \cite{Dicke1954}) which strongly couple with the guided modes in the waveguide. An effective model of this interface, which allows the nonreciprocal couplings \cite{Pichler2015}, can be written in Lindblad forms ($\hbar$ $=$ $1$), and the density matrix $\rho$ evolves as 
\bea
\frac{d \rho}{dt}=-i[H_L+H_R,\rho]+\mathcal{L}_L[ \rho]+\mathcal{L}_R[\rho],\label{rho}
\eea
where, respectively, the coherent and dissipative parts are
\bea
H_{L(R)} =&& -i\frac{\gamma_{L(R)}}{2} \sum_{\mu<(>)\nu}^N\left(e^{ik_s|r_\mu-r_\nu|} \sigma_\mu^\dag\sigma_\nu-\textrm{H.c.}\right)\label{LR}
\eea
and
\bea
\mathcal{L}_{L(R)}[\rho]=&&-\frac{\gamma_{L(R)}}{2} \sum_{\mu,\nu}^N e^{\mp ik_s(r_\mu-r_\nu)} \left(\sigma_\mu^\dag \sigma_\nu \rho + \rho \sigma_\mu^\dag\sigma_\nu \right.\nonumber\\
&&\left.-2\sigma_\nu \rho\sigma_\mu^\dag\right).
\eea

$\sigma_\mu^\dag$ $\equiv$ $|e\rangle_\mu\langle g|$ and $\sigma_\mu$ $=$ $(\sigma_\mu^\dag)^\dag$ are dipole operators, $k_s$ denotes the wave vector in the guided mode, and $\gamma_{L(R)}$ quantifies the left(right)-coupling rate. Equation (\ref{rho}) is obtained with Born-Markov approximation \cite{Lehmberg1970} in an interaction picture (energy difference $\omega_{eg}$ between the levels $|e\rangle$ and $|g\rangle$ is absorbed) under the reservoirs in the allowed dimension \cite{Tudela2013}. An intuitive and normalized directionality factor can be defined as \cite{Mitsch2014}
\bea
D=\frac{\gamma_R-\gamma_L}{\gamma_R+\gamma_L},
\eea
where $\gamma_R$ $+$ $\gamma_L$ $=$ $\Gamma$ $\equiv$ $2|dq(\omega)/d\omega|_{\omega=\omega_{eg}}g_{k_s}^2L$ \cite{Tudela2013} is the total decay rate. The inverse of group velocity is $|dq(\omega)/d\omega|$ with a resonant wave vector $q(\omega)$, the coupling strength is $g_{k_s}$, and the quantization length is $L$. A fractional $D$ $\in$ $[-1,1]$ quantifies the tendency and the amount of light exchange in the atomic array. For a periodic array of atoms with equal spacings, we use $\xi$ $\equiv$ $k_s |r_{\mu+1}-r_{\mu}|$ to characterize the strength of the resonant one-dimensional dipole-dipole interactions. We note that in Eq. (\ref{LR}), the atomic location has been ordered as $r_1$ $<$ $r_2$ $<...<$ $r_{N-1}$ $<$ $r_N$ for $N$ atoms.

{\it Bound and subradiant spin diffusion.}--When multiple atoms are excited initially, the system time dynamics can be solved directly from Eq. (\ref{rho}). Under the multi-atom excitations space, $|\phi_p\rangle$ $=$ $\sigma_j^\dag \sigma_{k>j}^\dag ...\sigma_{l}^\dag\sigma_{m>l}^\dag|0\rangle$ as the labeled $p$th bare state basis for $M$ excitations in general, the corresponding probability amplitudes $a_p(t)$ can be directly solved from
\bea
\dot{a}_p(t)=\sum_{q=1}^{C^N_M}V_{pq} a_q(t),
\eea   
where $V_{pq}$ denotes the matrix elements of the interaction kernel $V$ under a total of $C^N_M$ bare states with $C$ denoting the binomial coefficient. We further group these bare states into $(N-M+1)$ sectors \cite{SM, Jen2017_MP}, where each sector denotes one increment of the index of the first atomic excitation. Therefore, for example, there will be $(N-1)$ sectors for double excitation, where the first and the last bare states in the first sector should be $|e_1e_2g_3...g_N\rangle$ and $|e_1g_2g_3...g_{N-1}e_N\rangle$, respectively, while the last sector involves only one bare state $|g_1g_2...e_{N-1}e_N\rangle$. This is particularly useful for few atomic excitations space, where the excitation populations can be calculated in a systematic way and can be extended to a larger $M$ under a hierarchy relation \cite{SM}. For an even larger $M$ $\lesssim$ $N/2$, the cost of computation time increases exponentially as expected. From $a_p(t)$, we then obtain the excitation population $P_m(t)$ $=$ $\langle\psi(t)|\sigma^\dag_m\sigma_m|\psi(t)\rangle$ with $|\psi(t)\rangle$ $=$ $\sum_{p=1}^{C^N_M}a_p(t)|\phi_p\rangle$ \cite{SM}, and $\sum_{m=1}^N P_m(t)$ $=$ $M$ is the conserved quantity of total spin excitations.    

\begin{figure}[t]
\centering
\includegraphics[width=8.5cm,height=4.5cm]{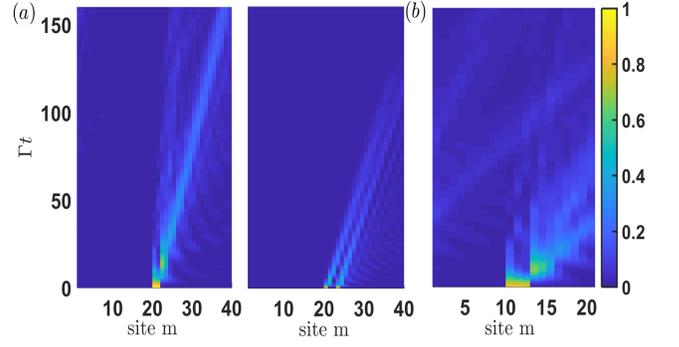}
\caption{Time evolutions of the bound dimers and trimers of atomic excitations in an atomic array. (a) Atomic excited state populations $P_m(t)$ evolve from two atomic excitations side-by-side (left panel) and separated by two lattice sites (right panel) at $D$ $=$ $0.1$ and $0.5$ respectively for $N$ $=$ $40$. (b) The time evolutions of the excitation populations $P_m(t)$ from three atomic excitations side-by-side at $D$ $=$ $0.1$ for $N$ $=$ $21$. In all plots, $\xi$ $=$ $\pi$.}\label{fig2}
\end{figure}

In Fig. \ref{fig2}, we plot the time dynamics of the doubly- and triply-excited state populations from the initialized product states $\sigma_j^\dag\sigma_k^\dag|0\rangle$ with $j-k=1$ and $2$ respectively in Fig. \ref{fig2}(a) and $\sigma_j^\dag\sigma_{j+1}^\dag\sigma_{j+2}^\dag|0\rangle$ in Fig. \ref{fig2}(b), with $j$ chosen near the center of the array. They propagate toward the end of the chain at long time and preserve their respective shapes. The long-time dynamics is particularly significant when $\xi$ is chosen close to $\pi$, where alternate minus and plus signs of hopping rates manifest in the excited atoms with a mutual separation of $(2n+1)\lambda/2$ and $n\lambda$ for an integer $n$, respectively. This subradiant dynamics has been investigated in a singly-excited atomic array \cite{Jen2020_subradiance}, and superradiance, by contrast, can show up when $\xi$ is close to $0$ or $2\pi$. 

The spin propagation is ballistic since its speed of spin diffusion is $\propto$ $D$ within each initialized configuration of the excited product states. For doubly-excited states, the bound spin excitations diffusion is faster when $|j-k|$ is smaller, which reflects the blockade of spin exchange between the nearest-neighbor atoms, effectively leading to a pushing force for the bound pair and the expedition of it at a shorter distance. The diffusion speed saturates as $|j-k|$ increases and approaches the single excitation limit as if there are no correlations in the independent atoms. For more spin excitations side-by-side in Fig. \ref{fig2}(b), an enhanced diffusion shows up but suffers from a larger intrinsic decay of $-M\Gamma/2$ for $M$ excitations. The interference patterns in Fig. \ref{fig2} is typical and common in the excitation propagation in the atom-waveguide interface \cite{Jen2020_disorder}, which results from multiple light reflections and transmissions throughout the atomic array.  

{\it Long-range and long-time correlations.}--To reveal the long-range and long-time characteristics of the multi-atom excitation transport, we employ the correlation functions via Kubo cumulant expansions \cite{Kubo1962}. Essentially the cumulant expansion reduces the mean values of the high order quantum correlations to the lower order cumulants of correlations. In particular, we take the average density-density and third-order correlation functions as the bulk properties of the atomic array \cite{Keesling2019}. They are, respectively,  
\bea
\langle G^{(2)}(r)\rangle\equiv&&\sum_j\frac{\langle n_j n_{j+r}\rangle-\langle n_j \rangle\langle n_{j+r}\rangle}{N-r},\\
\langle G^{(3)}\rangle\equiv&&\sum_j\frac{\langle n_j n_{j+1} n_{j+2}\rangle-\langle n_j \rangle\langle n_{j+1}\rangle\langle n_{j+2}\rangle}{N-2},
\eea 
where the site $r$ denotes the correlation length for doubly-excited spin diffusion. 

\begin{figure}[b]
\centering
\includegraphics[width=8.5cm,height=4.5cm]{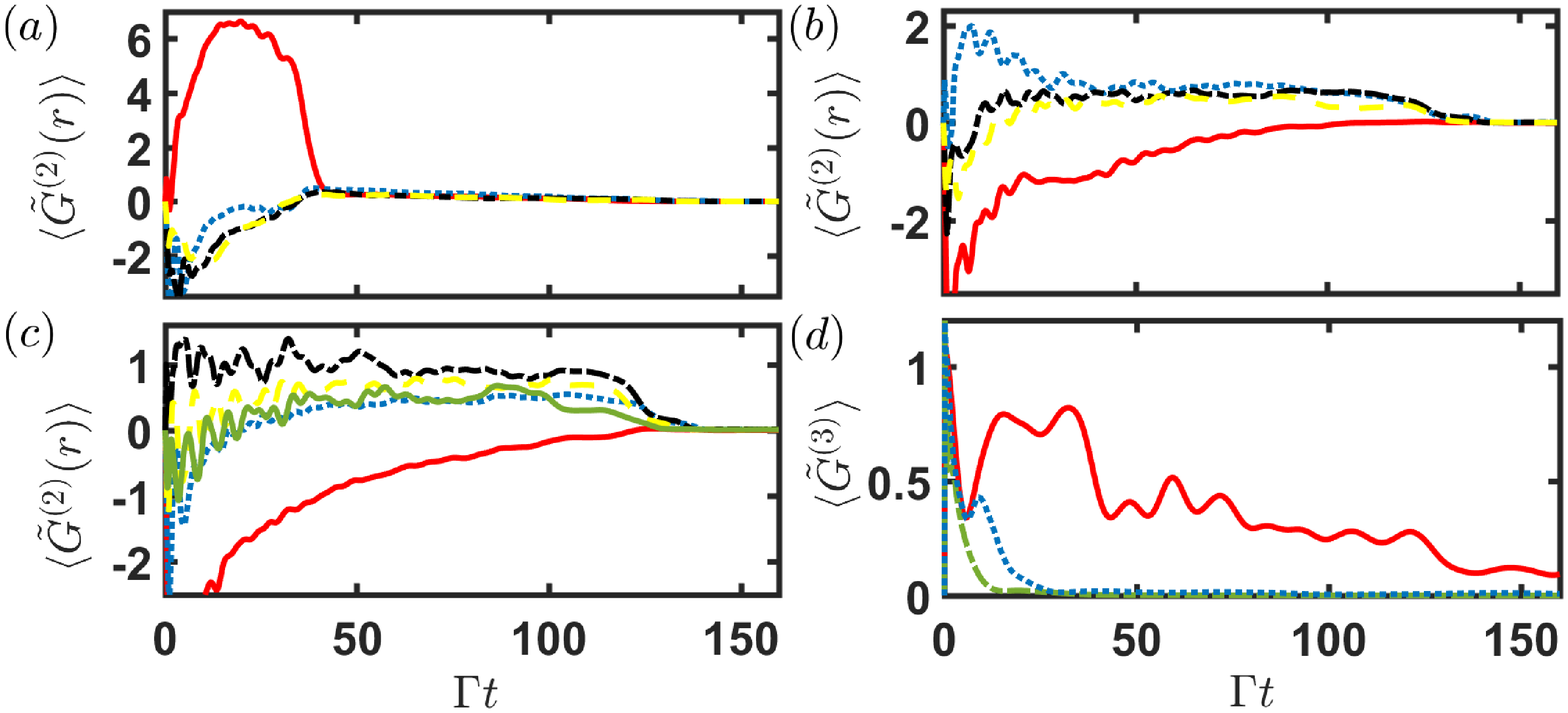}
\caption{Average density-density and third-order correlations. The scaled and time-dependent density-density correlations $\langle\tilde G^{(2)}(r)\rangle$ $=$ $\langle G^{(2)}(r)\rangle$ $\times$ $10^3$ arise at particular lattice separation $r$, when we initialize double atomic excitations at $r$ $=$ $1$ (solid line in red) in (a), $2$ (dotted line in blue) in (b), and $3$ (dash-dotted line in black) in (c), at $D$ $=$ $0.5$, $\xi$ $=$ $\pi$, and $N$ $=$ $40$. The correlation functions at longer $r$ $=$ $4$ (dashed line in yellow) and $5$ (solid line in green close to and right below the dashed line in yellow) are shown for comparison. In (d), the scaled third-order correlation $\langle\tilde G^{(3)}\rangle$ $=$ $\langle G^{(3)}\rangle$ $\times$ $10^2$ is shown for the case in Fig. \ref{fig2}(b) (solid line in red). As a comparison, the cases of $\xi$ $=$ $7.5\pi/8$ (dash-dotted line in green) and $7.8\pi/8$ (dotted line in blue) are shown.}\label{fig3}
\end{figure}

For the modified $\langle G^{(3)}\rangle$ we consider here as the third order correlation function side-by-side, it actually involves the third order cumulant and three other second order cumulants, which are \cite{Kubo1962} 
\bea
&&\left(\langle n_j n_{j+1}\rangle-\langle n_j \rangle\langle n_{j+1}\rangle\right)\langle n_{j+2}\rangle,\nonumber\\
&&\left(\langle n_j n_{j+2}\rangle-\langle n_j \rangle\langle n_{j+2}\rangle\right)\langle n_{j+1}\rangle,\nonumber\\
&&\left(\langle n_{j+1} n_{j+2}\rangle-\langle n_{j+1} \rangle\langle n_{j+2}\rangle\right)\langle n_{j}\rangle\nonumber.
\eea
Therefore, $\langle G^{(3)}\rangle$ represents the correlation function more than just the third order cumulant and can be regarded as the upper bound of it if the sum of these second order cumulants are positive.   

In Fig. \ref{fig3}, we numerically calculate the average and time-dependent density-density and modified third order correlation functions for various initialized product states of double excitations $\sigma_j^\dag\sigma_{j+r}^\dag|0\rangle$. Significant $\langle G^{(2)}(r)\rangle$ arises and sustains for long time at specific $r$, which highly correlates to the initialized excitation configuration. At a longer $r$ $>$ $1$, $\langle G^{(2)}(r'<r)\rangle$ is suppressed initially but revives at a later time. Interestingly, for most of the time dynamics before all correlations vanish, longer-range density-density correlations develop, and $\langle G^{(2)}(r'')\rangle$ $>$ $\langle G^{(2)}(r''')\rangle$ when $r'''>r''>r$. This suggests a sequential spread of the correlations when the spin excitations traverse through the atomic array, where $\langle G^{(2)}(r+1)\rangle$ grows preferentially over the others.  

The higher order correlation of the initialized product states of triple excitations $\sigma_j^\dag\sigma_{j+1}^\dag\sigma_{j+2}^\dag|0\rangle$ is shown in Fig. \ref{fig3}(d). Similar to the double excitations, a subradiant trimer of atomic excitations manifests in the coupling regime close to $\xi$ $=$ $\pi$. The $\langle G^{(3)}\rangle$ decays faster only slightly away from the subradiant coupling regime, indicating the fragility in maintaining the correlations in multiply-excited atomic excitations.  

{\it Encoded spin diffusion.}--Next we focus on the spin diffusion of the double excitations and propose an encoded spin dynamics depending on its initial coherences. Other than the initialized doubly-excited product states $\sigma_j^\dag\sigma_{j+1}^\dag|0\rangle$ with a null entanglement entropy $S$, we can initialize the system with a finite $S$. In general, the initial doubly-excited states that share the excitations separately can be expressed as
\bea
\left(\cos\phi \sigma_j^\dag+\sin\phi \sigma_{j+1}^\dag\right)\otimes\left(\cos\theta \sigma_{j+2}^\dag+\sin\theta \sigma_{j+3}^\dag\right)|0\rangle,\nonumber\\\label{theta}
\eea      
which features the initial $S$ $=$ $\ln 2$ and $2\ln 2$ for $(\phi,\theta)$ $=$ $(\pi/4,0)$ and $(\pi/4,\pi/4)$, respectively, at a cut right next to the sites $(j+2)$ and $(j+3)$. The most entangled doubly-excited state in the form of Eq. (\ref{theta}), on the other hand, should be equally distributed within the subspace,
\bea
&&\frac{1}{\sqrt{6}}\left(\sigma_j^\dag\sigma_{j+1}^\dag+ \sigma_j^\dag\sigma_{j+2}^\dag +\sigma_j^\dag\sigma_{j+3}^\dag +\sigma_{j+1}^\dag\sigma_{j+2}^\dag\right.\nonumber\\&&
\left.+\sigma_{j+1}^\dag\sigma_{j+3}^\dag+\sigma_{j+2}^\dag\sigma_{j+3}^\dag\right)|0\rangle,\nonumber
\eea
with $S$ $=$ $\ln 6$.

\begin{figure}[t]
\centering
\includegraphics[width=8.5cm,height=4.5cm]{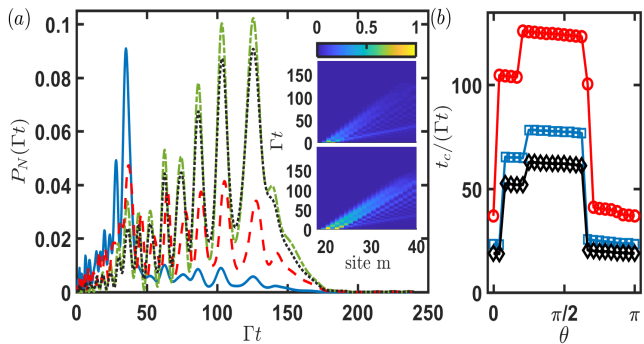}
\caption{Encoded spin diffusion. (a) At $D$ $=$ $0.5$, the end atomic excitation population probes the speed of spin diffusion for various initialized entanglement entropy $S$ $=$ $0$ (solid line in blue), $\ln 2$ (dashed line in red), $2\ln 2$ (dash-dotted line in green), and $\ln 6$ (dotted line in black). The insets present a faster (upper) and slower (lower) paired spin diffusion for the cases of $S$ $=$ $\ln 2$ and $2\ln 2$, respectively. (b) The $\theta$ dependence of the characteristic time $t_c$ for $D$ $=$ $0.5$ ($\circ$), $0.8$ ($\square$), and $1$ ($\diamond$). $\xi$ $=$ $\pi$ and $N$ $=$ $40$ are used in both plots.}\label{fig4}
\end{figure}

In Fig. \ref{fig4}(a), we plot $P_N(t)$ for four different initialized doubly-excited states as a probe of the spin diffusion at the edge of the array. They can be classified by the maximal $P_N(t_c)$ at which a small (large) $t_c$ corresponds to a null or lower (higher) $S$. This distinction of spin diffusion dynamics offers an opportunity for the state-dependent photon routing. 

In the insets of Fig. \ref{fig4}(a), the space-time dynamics of double spin diffusion presents different propagation speeds and interference patterns, where the respective maximal excitation population appears at the forefront and the tail in $P_N(t)$, respectively. The particular long period of the oscillations in $P_N(t)$ for $S$ $=$ $2\ln 2$ and $\ln 6$ before reaching their maximums can be seen as a precursor, in contrast to the case of a lower $S$ with afterglow fringes. These phenomena are also present in the case for a larger $D$ \cite{SM}. This suggests that more entangled initial excitations propagate more subradiantly close to the unidirectional coupling regime. 

For a smaller $D$ $\lesssim$ $0.5$, the interferences from the bi-directional light couplings are more involved in the spin diffusion dynamics, where the initially suppressed $\langle G^{(2)}(r)\rangle$ can revive and decline again, or even longer-range $\langle G^{(2)}(r)\rangle$ can emerge, making the encoded spin diffusion indistinguishable. For the purpose to demonstrate encoded spin diffusion, we focus on $D$ $\gtrsim$ $0.5$.       

From Eq. (\ref{theta}), we plot the the dependence of $t_c$ on $\theta$ at $\phi$ $=$ $\pi/4$ in Fig. \ref{fig4}(b). As expected, the time to reach the end of the atomic array is $\propto$ $D^{-1}$, and the maximal $t_c$ appears at $\theta$ $\approx$ $\pi/4$, corresponding to the most entangled state in our setting. The various $\theta$ dependence plateaus also indicate the robustness to the relative phase fluctuations in $\theta$. Therefore, a scheme using atom-waveguide interface to manipulate the state-dependent photon routing becomes feasible. 

{\it Discussion and conclusion.}--The atom-waveguide interface provides rich opportunities in processing quantum information \cite{Corzo2019} and quantum many-body simulations \cite{Kim2021}. The capability of this interface to reach the strong coupling regime \cite{Arcari2014, Tiecke2014, Yala2014} makes viable our prediction of the bound and subradiant dimers and trimers of atomic excitations. Although the subradiant spin diffusion is fragile to the position fluctuations of the periodic array, they can be mitigated by applying an optical lattice near the waveguide \cite{Corzo2019}. Moreover, it would be intriguing to further look into the dynamics of the shape-preserving multimers, which can benefit from long-time spin diffusion aided by the guided modes in the waveguide and thus may host quantum nonlinear interactions in photons \cite{Roy2017, Chang2018} with controlled strengths of dipole-dipole interactions and directionality of light couplings.   

In conclusion, we theoretically investigate the time-evolved density-density and third-order correlations in the atom-waveguide system from the initialized product or entangled states of multiple atomic excitations. Significant correlations can arise and sustain for long time, along with the shape-preserving characteristics in dimers and trimers of the excitations. We demonstrate that the encoded nonreciprocal spin diffusion is robust to their relative phase fluctuations, which can be useful and advantageous in quantum information processing, quantum storage and transport, and state-dependent photon routing.    

We acknowledge support from the Ministry of Science and Technology (MOST), Taiwan, under the Grant No. MOST-109-2112-M-001-035-MY3. We are also appreciated for insightful discussions with Jhih-Shih You and Ying-Cheng Chen. 



\clearpage
\appendix

\section{Construction of multi-atom excitations space}

The construction of multi-atom excitation space follows our previous work on multi-photon subradiant states \cite{Jen2017_MP}. The idea is simply that we first order N atomic positions as $r_1$ $<$ $r_2$ $<...<$ $r_{N-1}$ $<$ $r_N$, and then organize various excited sectors of the atomic excitations in sequential orders. 

We define the atomic labels $\vec\mu^n$ $\equiv$ $(\mu_1,\mu_2,...,\mu_M)^n$ for $M$ atomic excitations, where $n$ represents one specific order, and $\mu_j$ $\in$ $[j,N-M+j]$. Each order $n$ requires $\mu_j$ $\in$ $[\mu_{j-1}+1,N-M+j]$ with $\mu_j$ $<$ $\mu_{j+1}$. We then first increase $\mu_M$ $=$ $\mu_{M-1}$ $+$ $1$ up to $N$ while fix the other $\mu_j$'s. Next we keep increasing with an increment in $\mu_{M-1}$ $=$ $\mu_{M-2}$ $+$ $2$ along with $\mu_M$ $=$ $\mu_{M-1}$ $+$ $1$ up to $N$, until $\mu_1$ reaches ($N$ $-$ $M$ $+$ $1$). We take an example of $N$ $=$ $5$ and $M$ $=$ $2$, the bare states $|\psi_N^{(M)}(\vec\mu^n)\rangle$ are constructed and ordered as $|e_{1}e_{2}g_3g_4g_5\rangle$, $|e_{1}g_{2}e_3g_4g_5\rangle$, ..., $|e_{1}g_{2}g_3g_4e_5\rangle$, $|g_{1}e_{2}e_3g_4g_5\rangle$, $|g_{1}e_{2}g_3e_4g_5\rangle$, $|g_{1}e_{2}g_3g_4e_5\rangle$, $|g_{1}g_{2}e_3e_4g_5\rangle$,..., $|g_{1}g_{2}g_3e_4e_5\rangle$. This leads to a total of $20$ states and $C^N_M$ of them in general with a binomial coefficient $C$. 

Next we obtain the coupled equations of the probability amplitudes according to Eq. ($1$) in the main paper, 
\bea
\dot{a}_p(t)=\sum_{m=1}^{C^N_M} V_{pq}a_q(t),
\eea
where $V_{pq}$ couples all $C^N_M$ states in the subspace. The matrix elements are $V_{pp}$ $=$ $-\frac{M\Gamma}{2}$, and $V_{p,q\neq p}$ $=$ $(-F_{s_1s_2}+iG_{s_1s_2})$ where two indices $(s_1,s_2)$ can be obtained from a sorting function $Sort(p,q)$ we implement in the software $Matlab$. The general forms of $V_{p,q}$ can be obtained as
\begin{widetext}
\bea
V=&&\begin{bmatrix}
    -F_{11}  & -F_{12}+iG_{12} 		& -F_{13}+iG_{13} & \dots & -F_{1N}+iG_{1N} \\
    -F_{12}^*+iG_{12}^* & -F_{22} & -F_{23}+iG_{23}	& \dots & -F_{2N}+iG_{2N} \\
    -F_{13}^*+iG_{13}^* & -F_{23}^*+iG_{23}^* & -F_{33} & \dots & -F_{3N}+iG_{3N} \\
		\vdots 							& \vdots  & \vdots & \ddots & \vdots  \\
		-F_{1N}^*+iG_{1N}^* & -F_{2N}^*+iG_{2N}^* & -F_{3N}^*+iG_{3N}^* & \dots &-F_{NN}
\end{bmatrix},
\eea
\end{widetext}
where 
\bea
F_{\mu\nu}\equiv&&\frac{\gamma_Re^{ik_s|r_{\mu,\nu}|}+\gamma_Le^{-ik_s|r_{\mu,\nu}|}}{2},\\
G_{\mu\nu}\equiv&&-i\frac{\gamma_Re^{ik_s|r_{\mu,\nu}|}-\gamma_Le^{-ik_s|r_{\mu,\nu}|}}{2},
\eea
and $r_{\mu,\nu}$ $\equiv$ $r_\mu-r_\nu$. We can further reduce the above to 
\begin{widetext}
\bea
V=&&\begin{bmatrix}
    -\frac{\gamma_L+\gamma_R}{2}  & -\gamma_Le^{-ik_s|r_{1,2}|} & -\gamma_Le^{-ik_s|r_{1,3}|} & \dots & -\gamma_Le^{-ik_s|r_{1,N}|}\\
    -\gamma_Re^{-ik_s|r_{1,2}|} & -\frac{\gamma_L+\gamma_R}{2} & -\gamma_Le^{-ik_s|r_{2,3}|} & \dots & -\gamma_Le^{-ik_s|r_{2,N}|}\\
    -\gamma_Re^{-ik_s|r_{1,3}|} & -\gamma_Re^{-ik_s|r_{2,3}|} & -\frac{\gamma_L+\gamma_R}{2} & \dots & -\gamma_Le^{-ik_s|x_{3,N}|}\\
		\vdots 							& \vdots  & \vdots & \ddots & \vdots  \\
		-\gamma_Re^{-ik_s|r_{1,N}|} & -\gamma_Re^{-ik_s|r_{2,N}|} & -\gamma_Re^{-ik_s|r_{3,N}|} & \dots & -\frac{\gamma_L+\gamma_R}{2}
\end{bmatrix},
\eea
\end{widetext}
which has nonsymmetric feature of the nonreciprocal coupling matrix.

We use $Sort$ to sort out two numbers, $s_1$ and $s_2$, after comparing the $p$th and $q$th bare states of $|\psi^{(M)}_N(\vec{\mu})\rangle$, which correspond to one different excited atomic index in these bare states, respectively. We take the example of $N$ $=$ $4$ and $M$ $=$ $2$. To determine $V_{1,2}$, $Sort(1,2)$ $=$ $Sort(|e_{1}e_{2}g_3g_4\rangle,|e_{1}g_{2}e_3g_4\rangle)$ gives $(2,3)$, showing that the dipole-dipole interaction lowers the third atomic excited state while raising the second atom to the excited one. If $(s_1,s_2)$ gives $(0,0)$, it leaves a null $V_{n,m\neq n}$ when more than one distinct atomic indices appear in the $p$th or $q$th bare states. Using the same example, $V_{1,6}$ ($V_{2,5}$) $=$ $0$, since there is no dipole-dipole interaction coupling between the states $|e_{1}e_{2}g_3g_4\rangle$ and $|g_{1}g_{2}e_3e_4\rangle$ ($|e_{1}g_{2}e_3g_4\rangle$ and $|g_{1}e_{2}g_3e_4\rangle$) \cite{Jen2017_MP}.

\section{Derivations of atomic excitation populations}

To calculate the atomic excitation population $P_m(t)$, we group the bare states introduced above into $(N-M+1)$ sectors, where each sector denotes one increment in the label $\mu_1$. Therefore, there will be $(N-1)$ sectors for double excitations. The first bare state in the first sector is $|e_1e_2g_3...g_N\rangle$, while the last sector involves only one bare state $|g_1g_2...e_{N-1}e_N\rangle$. This is particularly useful for few atomic excitations space, where the excitation populations can be calculated in a systematic way and can be extended to a larger $M$ under a hierarchy relation.

Within the $n$th sector, we use the labels $s^{(M)}_n$ to denote $\vec\mu^n$ in the above, where $s^{(M)}_n(l)$ denotes the $l$th bare state basis in this sector, and $dim(s^{(M)}_n)$ denotes the total number of the elements in the sector. The number of elements in the first sector $s$ for $M$ excitations is denoted as $N_{s,1}^{(M)}$ $=$ $C^{N-1}_{M-1}$. Within each $n$th sector, we use the label $N^{(M)}_{s,n}$ $=$ $N^{(M)}_{s,n-1}+dim(s^{(M)}_n)$ as the total number of elements up to the $n$th sector. We also define $N^{(1)}_{s,\alpha}$ $=$ $1$, $N^{(\alpha)}_{s,0}$ $=$ $0$ for arbitrary $\alpha<N$. 

The final results for general atomic excitation populations $P_m^{(M)}(t)$ are
\begin{widetext}
\bea
P_m^{(M)}(t)=\sum_{n=s^{(M)}_m(1)}^{s^{(M)}_m[dim(s^{(M)}_m)]}|a_n(t)|^2+\delta_{M>2}\sum_{k=1}^{M-2}~\sum_{n=1}^{m-k>0}\sum_{l=l_I}|a_l|^2+\left(\delta_{M=2}\delta_{k=1}+\delta_{M>2}\sum_{k=M-1}^{N-1}\right)\sum_{n=1}^{m-k>0}|a_{l''}|^2
\eea 
\end{widetext}
where $l_I$ $=$ $1+N^{(M)}_{s,n-1}+\sum_{l'=k}^{m-n-1}(N-m+1+l')^{(M-1)}_{s,1}$ and $l''$ $=$ $k-(M-2)+N^{(M)}_{s,n-1}+\sum_{l'=k}^{m-n-1}(N-m+1+l')^{(M-1)}_{s,1}$. The Kronecker delta function is $\delta$. The hierarchy relation is embedded in the sums of index $k$, which becomes cumbersome as $M$ increases owing to the large Hilbert space of a total $C^N_M$ of them.    

\section{Encoded spin diffusion for a larger $D$}

\begin{figure}[bh]
\centering
\includegraphics[width=8.5cm,height=4.5cm]{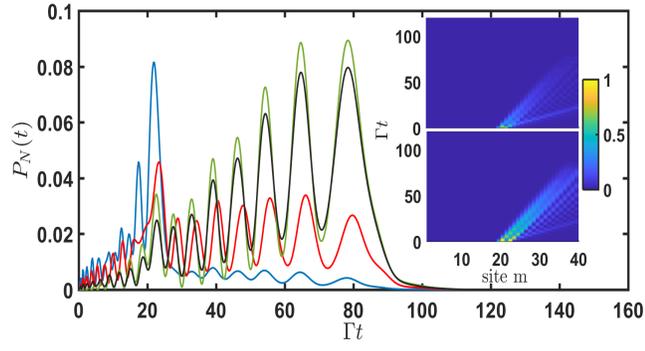}
\caption{Encoded spin diffusion for a larger $D$ $=$ $0.8$. The atomic excitation population $P_N(t)$ probes the speed of spin diffusion for various initialized entanglement entropy $S$ $=$ $0$ (solid line in blue), $\ln 2$ (solid line in red), $2\ln 2$ (solid line in green), and $\ln 6$ (solid line in black). The insets present a faster (upper) and slower (lower) spin diffusion for the cases of $S$ $=$ $\ln 2$ and $2\ln 2$, respectively. $\xi$ $=$ $\pi$ and $N$ $=$ $40$ are used here.}\label{figs1}
\end{figure}

Here we demonstrate the encoded spin diffusion for a different and a larger $D$ in Fig. \ref{figs1}. The case for a larger $D$ indicates a faster spin propagation in general. Similar observation is shown here, where the initialized double excitation with a null or lower entanglement entropy $S$ propagates faster than the one with a higher $S$. This indicates a distinction of spin diffusion from different initialized coherence properties of the states and suggests a potential application to state-dependent photon routing. 

\end{document}